a

\documentclass[12pt]{iopart}
\usepackage{graphicx}
\usepackage{verbatim}
\usepackage{subfigure}
\usepackage{float}
\restylefloat{figure}

\begin{document}

\title{Optical identification using imperfections in 2D materials}

\author{Yameng Cao$^1$, Alexander J Robson$^1$, Abdullah Alharbi$^{2}$, Jonathan Roberts$^{1}$, Christopher S Woodhead$^{1}$, Yasir J Noori$^{1}$, Ram\'{o}n Bernardo-Gavito$^{1}$, Davood Shahrjerdi$^{2}$, Utz Roedig$^{3}$, Vladimir I Fal'ko$^{4}$ and Robert J Young$^1$}

\address{$^1$ Physics Department, Lancaster University, Lancaster, LA1 4YB, UK}
\address{$^2$ Electrical and Computer Engineering, New York University, Brooklyn, New York 11201, USA}
\address{$^3$ School of Computing and Communications, Lancaster University, Lancaster, LA1 4WA, UK}
\address{$^4$ National Graphene Institute, The University of Manchester, Manchester, M13 9PL, UK}

\ead{r.j.young@lancaster.ac.uk and y.cao5@lancaster.ac.uk}
\vspace{10pt}
\begin{indented}
\item[]July 2017
\item[]Keywords: Physical unclonable function, TMD, optical measurement, photoluminescence, security
\end{indented}

\vspace{10pt}

\begin{abstract}
The ability to uniquely identify an object or device is important for authentication \cite{Roberts2015}. Imperfections, locked into structures during fabrication, can be used to provide a fingerprint that is challenging to reproduce. In this paper, we propose a simple optical technique to read unique information from nanometer-scale defects in 2D materials. Imperfections created during crystal growth or fabrication lead to spatial variations in the bandgap of 2D materials that can be characterized through photoluminescence measurements. We show a simple setup involving an angle-adjustable transmission filter, simple optics and a CCD camera can capture spatially-dependent photoluminescence to produce complex maps of unique information from 2D monolayers. Atomic force microscopy is used to verify the origin of the optical signature measured, demonstrating that it results from nanometer-scale imperfections. This solution to optical identification with 2D materials could be employed as a robust security measure to prevent counterfeiting.
\end{abstract}

\section{Introduction}
Physically unclonable functions (PUFs) provide a method to generate secrets for unique identification or cryptographic key generation \cite{Gao2016}. Instead of storing the secret in digital memory, or asking a user to provide it, it is derived from a physical characteristic of the system. A PUF can be constructed in various ways, including scattering patterns of an optical medium \cite{Pappu2002} or chip-specific transistor switch delay variations \cite{Anderson2008}. The assumption is that the secret cannot be copied, as it is bound to a physical entity which cannot be cloned. Furthermore, it is assumed that the probability of finding two devices with identical physical characteristics is very low.

Existing PUFs have limitations, as they are often difficult to produce, and more importantly, there is no guarantee that they cannot be cloned. Arbiter PUFs, Ring Oscillator PUFs, XOR PUFs, Lightweight Secure PUFs and Feed-Forward PUFs have all been attacked using machine learning techniques \cite{Ruhrmair2010}. To address the shortcomings, the research community has looked at variations in nanoscale devices \cite{Gao2016}. These include the use of memristors \cite{Gao2015}, fabrication variability in magnetic random access memories \cite{Vatajelu2016}, unique characteristics in carbon nanotube transistors \cite{Konigsmark2014} and phase change memories \cite{Zhang2013}. These solutions vary in the practicality of implementation, however, they are not sensitive to the smallest imperfections at the atomic scale. This is important because at the lower limit of physical size, cloning of a physical entity requires identifying the chemical makeup as well as dealing with the probabilistic nature of quantum mechanics. Atomic scale imperfections, such as defects in a crystal lattice represent this category of entities. In 2D materials, vacancies, impurities, grain boundaries and other structural defects, lead to spatially varying contributions from excitonic complexes that complicate the photoluminescence. Though spatial non-uniformity could prove detrimental to optoelectronics, the fact that these variations originate from atomic level defects is an advantage for implementing unique optical identifiers, using transition metal dichalcogenides (TMDs) monolayers as optically varying physical unclonable functions (OPUFs).

Spatial inhomogeneity in the photoluminescence (PL) of TMD monolayers, depend on the method of synthesis. Mechanically exfoliated flakes exhibit the highest uniformity while flakes grown by chemical vapor or physical vapor transport are non-uniform in general, due to impurities and defect induced doping. In this work, we demonstrate a method to extract information from the photoluminescence of tungsten disulphide WS$_{2}$ suitable for unique identification or authentication.

\section{Method}
\begin{figure}[H]
\centering
\includegraphics[width = 150mm]{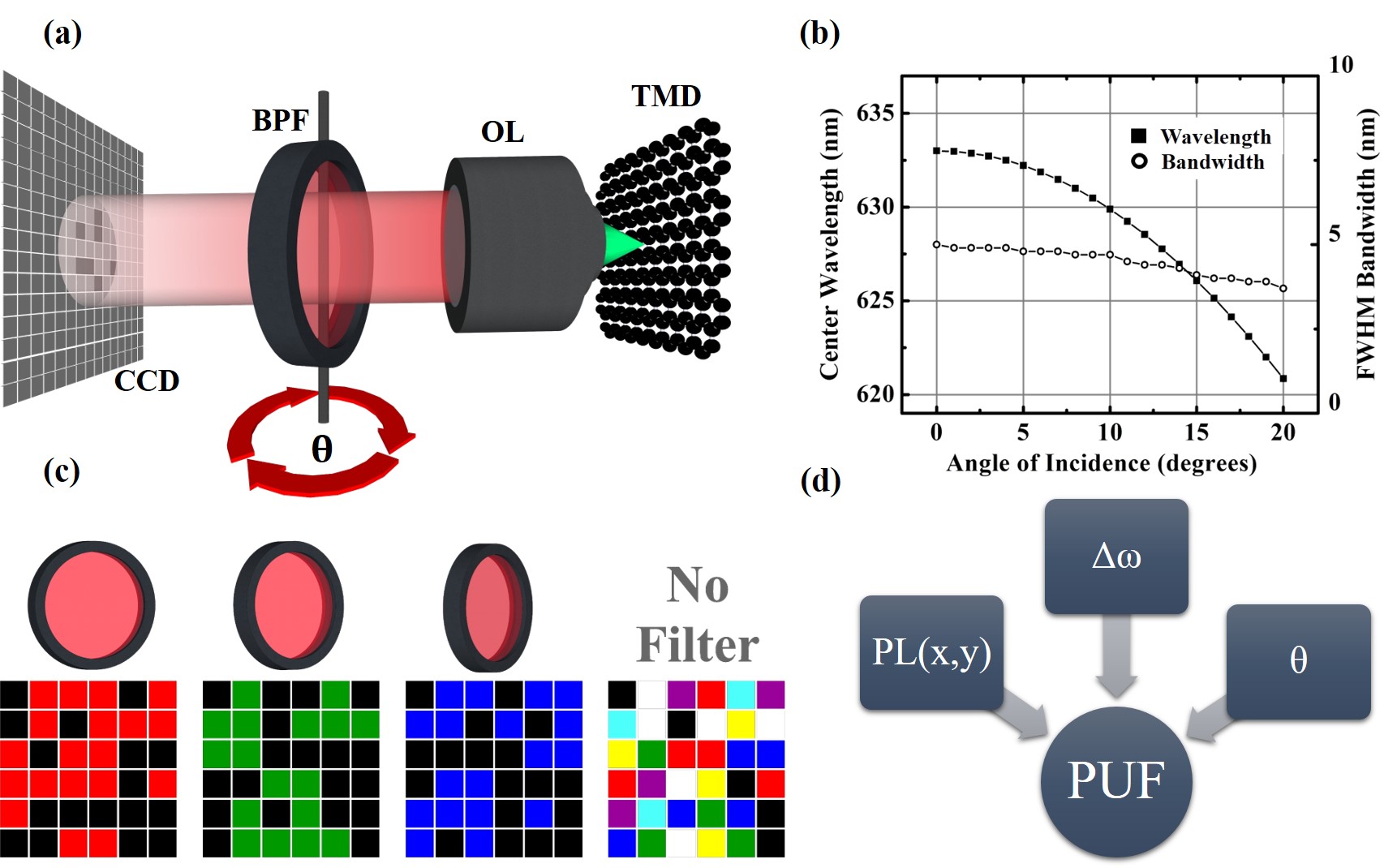}
\caption{Extracting unique information from 2D materials. (a): Measurement apparatus, in which the photoluminescence from a monolayer TMDC is collected by an objective lens (OL), selectively transmitted through a rotatable optical bandpass filter (BPF), finally imaged on a CCD sensor. Angular orientations of the BPF determines the center-wavelength of its pass band, which varies with incidence angle as shown in (b). (c): Concept of the angular selective transmission. Changing the BPF angle lights up a random subset of pixels on the CCD; red, green and blue conceptually correspond to positions on the monolayer TMDC that emits in differing energy ranges. When no filter is present, all energies are picked up. (d): The BPF angular orientation $\theta$, the corresponding BPF bandwidth $\Delta\omega$, and the spatially varying photoluminescence of the monolayer TMDC PL(x,y) makes up the physical unclonable function.}
\label{Fig1}
\end{figure}

Our concept of extracting unique information from the bandgap of 2D materials is remarkably simple, as depicted in Figure \ref{Fig1}(a). A lens collects the photoluminescence from a monolayer TMD, a rotatable optical band-pass filter (BPF) selects a wavelength-range of emitted photons based on its angle dependent transmission wavelength $\lambda(\theta)$ and bandwidth $\Delta\omega(\theta)$, then the transmission is mapped by a charge coupled device (CCD). Direct band gap single atomic layers of the TMD family have shown tremendous potential for applications in optoelectronics \cite{Wu2015,Geim2013}, quantum photonics \cite{Tran2015,Ye2015}, spin and valley spin physics \cite{Wu2016}. However, to extract unique information efficiently for optical security applications \cite{Pappu2002}, high internal quantum efficiency $\eta_{q}$ would be required to maximize the rate of authentication. Various demonstrations \cite{Gan2013,Wu2014,Amani2015,Wang2016,Alharbi2017} have shown that it is possible to improve $\eta_{q}$ especially at room temperature, which is limited by non-radiative processes. For the scope of this paper, we chose to study WS$_{2}$ for its high $\eta_{q}$ at room temperature \cite{Mak2016} without enhancement. A light source, for instance a laser or a lamp illuminates a region of the sample containing monolayer WS$_{2}$, generating photoluminescence via above-band excitation.

We define the angle of incidence $\theta$ ordinarily as the angle which the transmission from the lens makes with the normal to the filter interface. The angle dependent properties of the BPF is illustrated in Figure \ref{Fig1}(b), where we firstly find a blueshift in the transmission wavelength $\lambda$ then also a narrowing of the full-width at half-maximum transmission bandwidth $\Delta\omega$, as the angle of incidence is varied. This stems from the fact that at the inter-layer interfaces of dielectric interference filters, the path difference between transmitted and reflected rays reduce with increasing angle of incidence. Correspondingly, the wavelength shift may be described by the following well-known approximation,
\begin{center}
  \begin{equation}
  \lambda=\lambda_{0}\sqrt{1-\frac{\sin^{2}\theta}{n^{2}}} \label{eqnblueshift},
\end{equation}
\end{center}

where $\lambda_{0}$ and $n$ denotes the zero-angle transmission wavelength and effective refractive index, respectively \cite{Baumeister2014}. Transmission measurements of the BPF, manufactured by Thorlabs was not known, though with the transfer matrix method we simulated a generic distributed Bragg structure with similar transmission characteristics as the BPF used in our apparatus. From this simulation we extracted the change in bandwidth with the angle of incidence, as plotted in Figure \ref{Fig1}(b).

Changing the angle of incidence selects a specific region on the emission spectrum to be detected by the CCD. Illustratively, colored squares in Figure \ref{Fig1}(c) represent light-sensitive pixels on a CCD corresponding to a specific energy range according to $\lambda(\theta)$ and $\Delta\omega(\theta)$. When $\theta=0$, transmission matches the filter specification, but for non-zero angles the transmission blueshifts hence green and eventually blue are being detected. In the absence of any bandpass filter and specular reflections due to the excitation source, the camera records a superposition of all energies for each spatial co-ordinate $(x,y)$. Hence pixels are white if photons of all energies arrive, purple if only red and blue arrive, etc. It may seem possible to take the no-filter image and algorithmically extract the red, green and blue constituents of the emission spectrum, enabling reverse engineering of $\lambda(\theta)$. However in practice, for the monolayers we have tested in this work, the bandgap variation in wavelength is on the order of a few nanometers, or tens of mili-electronvolt in energy, i.e. sufficiently small that the color filters in a CCD would not be able to differentiate between the emission energies.

We prepared samples of WS$_{2}$ using mechanical exfoliation as well as chemical vapor deposition (CVD). The exfoliated samples were made using Nitto water soluble tape hosting the bulk crystals, which were adhered to a 0.5 mm thick polydimethylsiloxane (PDMS) layer from Gel-Pak on a glass slide, then quickly peeled off. Individual flakes were optically identified and their thickness spectrally confirmed with both room temperature photoluminescence and Raman optical measurements. Chemical vapor deposition was carried out at 950$^{\circ}$C on $p^{+}$ silicon substrates with 285 nm thermally grown SiO$_{2}$, using optimized chamber conditions that allowed large monolayer flakes to be obtained \cite{Alharbi2016}.

\section{Results}
\subsection{WS$_{2}$ from Mechanically Exfoliation}
Defects can degrade the optical quality of monolayer emitters, potentially harming the efficiency of optoelectronic devices. For authentication purposes as a PUF, however, atomic level disorder is a bonus. Defects that tend to increase the spatial inhomogeneity in photoluminescence, the distribution, density and variety of which all contribute to the structural complexity of the optical PUF. We found that the variation in emission energy with filter angle is more pronounced with CVD flakes, though sufficiently large area exfoliated flakes may also be suitable for this application.

\begin{figure}[H]
\centering
\includegraphics[width = 150mm]{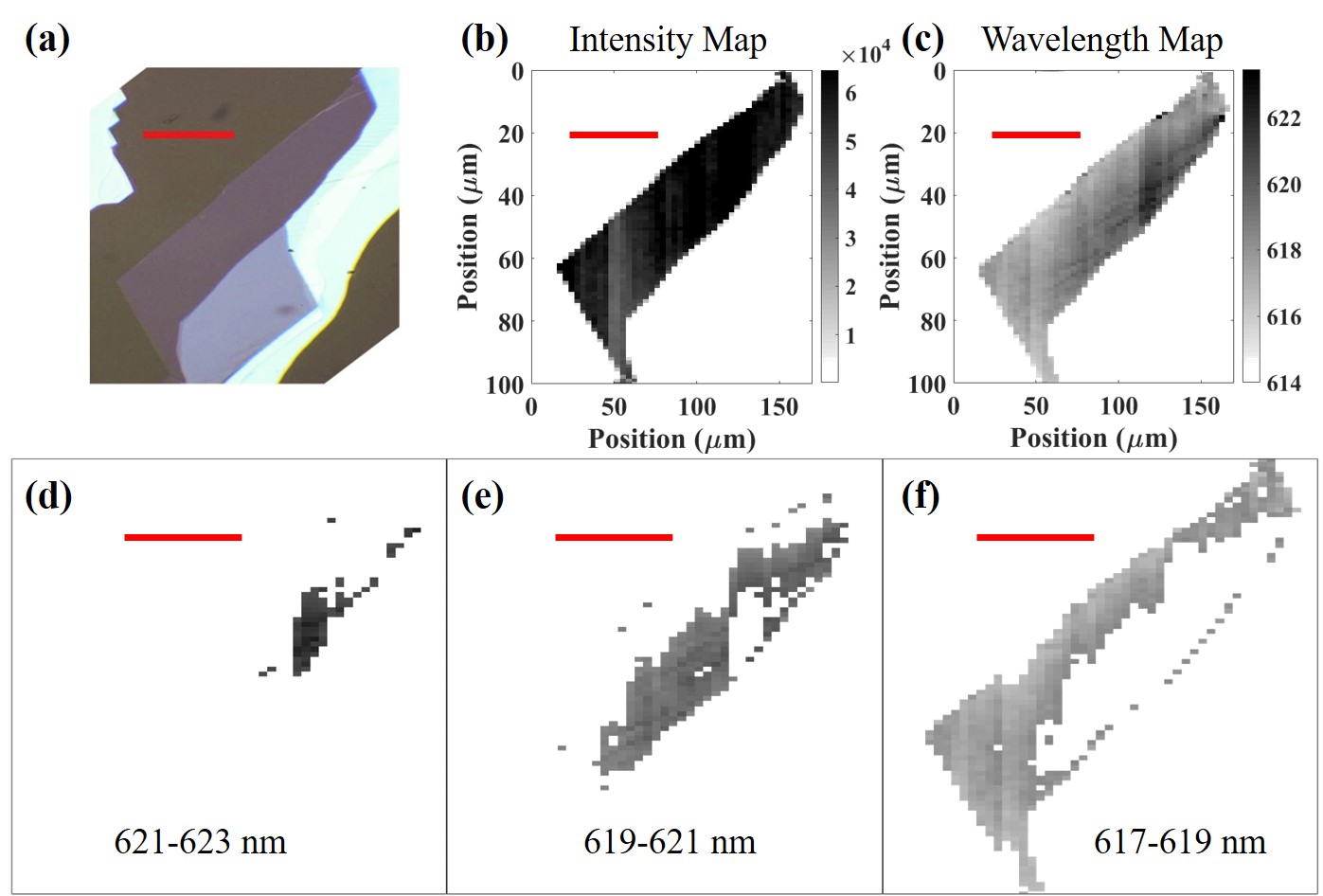}
\caption{(a): 50x Optical image of the exfoliated WS$_{2}$ flake on PDMS. $\mu$-PL map of this flake was recorded with 532 nm excitation and 100 \textrm{$\mu$}W excitation power at 300 K. The integration time for each pixel is 0.5 seconds. We took spectra from each map co-ordinate and used a Gaussian fit to determine the peak intensity (in counts per second) and the peak wavelength (in nm), plotted in (b) and (c) respectively. In (d), (e) and (f) we applied a rectangular function that selected pixels with peak wavelength falling within a certain band. In all cases the scale bar corresponds to 50 $\mu$m.}
\label{Fig2}
\end{figure}

In Figure \ref{Fig2}, we excite an exfoliated monolayer WS$_{2}$ on PDMS with 532 nm laser, focused to a sub-micron spot with a 0.9 NA objective lens. The same objective collects the PL, which was then measured by a Horiba LabRAM HR spectrometer. PDMS, being an elastomer, relaxes strain in the monolayer, enabling extraordinarily large flakes \textit{hundreds} of microns in length to be routinely obtained, as shown by figure \ref{Fig2}(a). In comparison to recent work on large area TMD mechanical exfoliation \cite{Magda2015}, this method does not require an atomically flat gold substrate and eliminates the intermediate sonication step. Here, the $\mu$-PL measurement serves the purpose of showing the photoluminescence inhomogeneity across the samples, this method is impractical for real PUF implementation however, due to it's size and cost. We used Gaussian-type fitting for each pixel and extracted maps of peak intensity and wavelength, respectively shown in Figure \ref{Fig2}(b) and (c).

Next, we applied a rectangular function that zeroed the value of all pixels except those with a peak wavelength falling within a specific range, approximating the effect of an ideal optical bandpass filter. Changing this filter wavelength range then simulates different angle of incidence, as we described previously. This simple function highlights the PL spatial variation within a monolayer and by changing the detection spectral window, corresponding to $\theta$ in Figure \ref{Fig1}(a), different spatial regions can be turned on and off. We showed three spectral windows in Figure \ref{Fig2}(d)-(f), corresponding to increasing the BPF incidence angle. This simple function takes into account the peak wavelength as well as the gradient of spatially varying intensity, but it exaggerates the differences between each filter orientation for visual effect. In practice, the difference from one angle to the next is less obvious, since the transmission band picks out a range of wavelengths, rather than only those with peak wavelength in the window.

The most common defects in exfoliated disulphide TMD is zero-dimensional sulfur vacancies \cite{Zhou2013}. Upon the removal of a sulphur atom, the transition metal is left with an unpaired electron, which introduces unintentional n-doping \cite{McDonnell2014}. On the other hand, W or Mo vacancies could lead to unintentional p-doping. Bandgap renormalization occurs as a result of the doping, as confirmed by the shift in peak emission wavelength across the sample, depicted in Figure \ref{Fig2}(c). Adatoms can also lead to similar doping effects, as demonstrated by several studies \cite{Chen2017,Tedstone2016}.

\subsection{WS$_{2}$ from Chemical Vapor Deposition}
The CVD growth mode is based on a bottom-up crystal nucleation on a target substrate. While other bottom-up synthesis methods exist such as atomic layer deposition \cite{Tan2014}, pulsed laser deposition \cite{Late2014} and physical vapor deposition \cite{Liu2014}, crystal size and the quality of the flakes are limited in general compared to CVD. Quality of flakes grown using CVD is heavily dependent on reaction kinetics, which is varied and lead to immense difficulty in reproducing the same growth conditions across different reactors. Monolayers synthesized by chemical vapor growth have defective properties that depend on the types of substrate being used, lattice mismatch at the material-substrate interface, thermal stability and chamber conditions. In addition to point defects, CVD grown monolayer TMDs also exhibit one-dimensional defects, such as agglomerations of sulfur vacancies in a line \cite{Han2015,Komsa2013}, grain boundaries \cite{Lin2015,Lehtinen2015} and dislocations \cite{Azizi2014,Zou2013}.

\begin{figure}[H]
\centering
\includegraphics[width = 150mm]{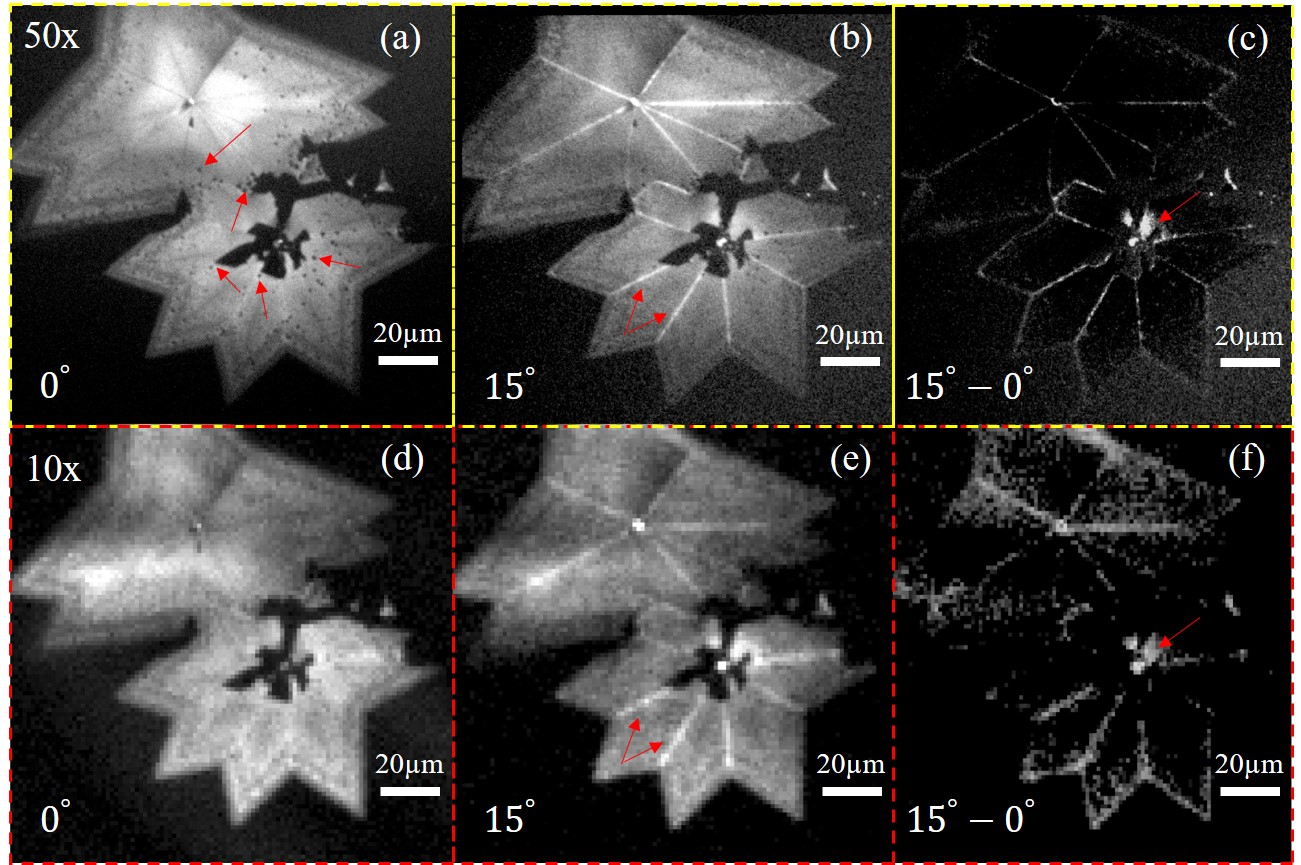}
\caption{Angular-dependent PL images of WS$_{2}$ monolayer flake, excited by 450 nm laser, collected using 50x (a)-(c) and 10x (d)-(f) respectively. From left to right: $0^{\circ}$ with arrows indicating dark spots, $15^{\circ}$ with arrows indicating bright lines, then intensity differences between $0^{\circ}$ and $15^{\circ}$ with arrows indicating the area around nucleation. }
\label{Fig3}
\end{figure}

Figure \ref{Fig3} shows the images recorded using filter-angle modulation, for a selected WS$_{2}$ flake. In each case, PL was imaged through a bandpass filter, Thorlabs FLH633-5, center wavelength $633$ nm and bandwidth $5$ nm, attached to an angle-variable mount, following illumination using a 450 nm laser at an angle. Laser power density is approximately $3$ Wcm$^{-2}$. In this arrangement, the microscope effectively operates in darkfield mode. The objective lenses used were an Olympus long working distance 50x and a Zeiss 10x. The 10x images were deliberately cropped to highlight the same flake as recorded using the 50x objective lens. Lastly the CCD is a thermoelectrically cooled Sony ICX825 sensor array.

Once $\theta$ is set to $0^{\circ}$, we focused the image under brightfield white light illumination first, then with only laser illumination an image is acquired. As the BPF transmits a narrow band centered at $633$ nm at $\theta=0$, only the single layer WS$_{2}$ photoluminescence is imaged. Subsequent images for nonzero $\theta$ were acquired in a similar fashion. Due to undesired reflections within the filter, the transmitted PL also acquired a slight deflection before reaching the sensor, for changing $\theta$. This deflection translates into a lateral image shift. After image acquisition, we processed the image series using MATLAB in the order of image registration, background subtraction and pixel-level analysis. Image registration matches geometric features from an image to the next, moving each laterally so that the same features align in all images, correcting the image shift. For background subtraction, a rectangular area representative of the background in each image was selected, averaged over all pixels in the selection and subtracted the average from all other pixels in the original image. We found that this is equivalent to dark frame subtraction in effect, but much easier to implement. The analysis carried out is as follows. First the greyscale value at each pixel represented by an $(x,y)$ co-ordinate for each $\theta$ is extracted, into a four-dimensional matrix $I(x,y,\theta)$, where $I$ denotes the pixel value.  Then, for each $I(x,y,\theta)$ image in the series, $I(x,y,\theta_{0}),I(x,y,\theta_{1},...)$, the difference $I(x,y,\theta_{n})-I(x,y,\theta_{0})$ was found, where $n$ is the index of an image in the sequence. We then repeated the entire procedure from acquisition to analysis for the 10x objective lens.

The CVD grown domains are a mixture of triangular and hexagonal shapes. Different geometries in growth arise from variations in the growth rate of different edge terminations. In all cases, the domains start with a nucleation site, growth then favors the sides of the nucleation with the highest surface free energy, which leads to different final domain shapes. Figure \ref{Fig3} shows a chosen large area domain. As $\theta$ is increased, a number of visually striking features can be observed. Firstly, lines connecting inner vertices and the nucleation site become brighter, they are likely to be grain boundaries between edges with differing terminating atoms. Then, for low angles we can see a distribution of ``spots'', which vanish for higher angles. The emission intensity for the entire domain attenuates in general for higher angles, but the opposite is true around the nucleation site. Lastly, a dark concentric band can be seen that does not change with $\theta$. The fact that these emissive phenomena are picked out by the BPF at higher angles suggest a blueshift in the optical spectrum, possibly caused by higher order of quantum confinement. Their origins are unknown to us at present and will warrant further investigation. Though already visually clear, the pixel difference maps in Figure \ref{Fig3} highlight the effect of filter angle modulation, showing the blueshifted emission from surface defects. The fact that we can image these defects even with a low magnification objective lens implies that our system for unique information extraction can potentially be simplified into one with much lower overhead.

\begin{figure}[H]
\centering
\includegraphics[width = 150mm]{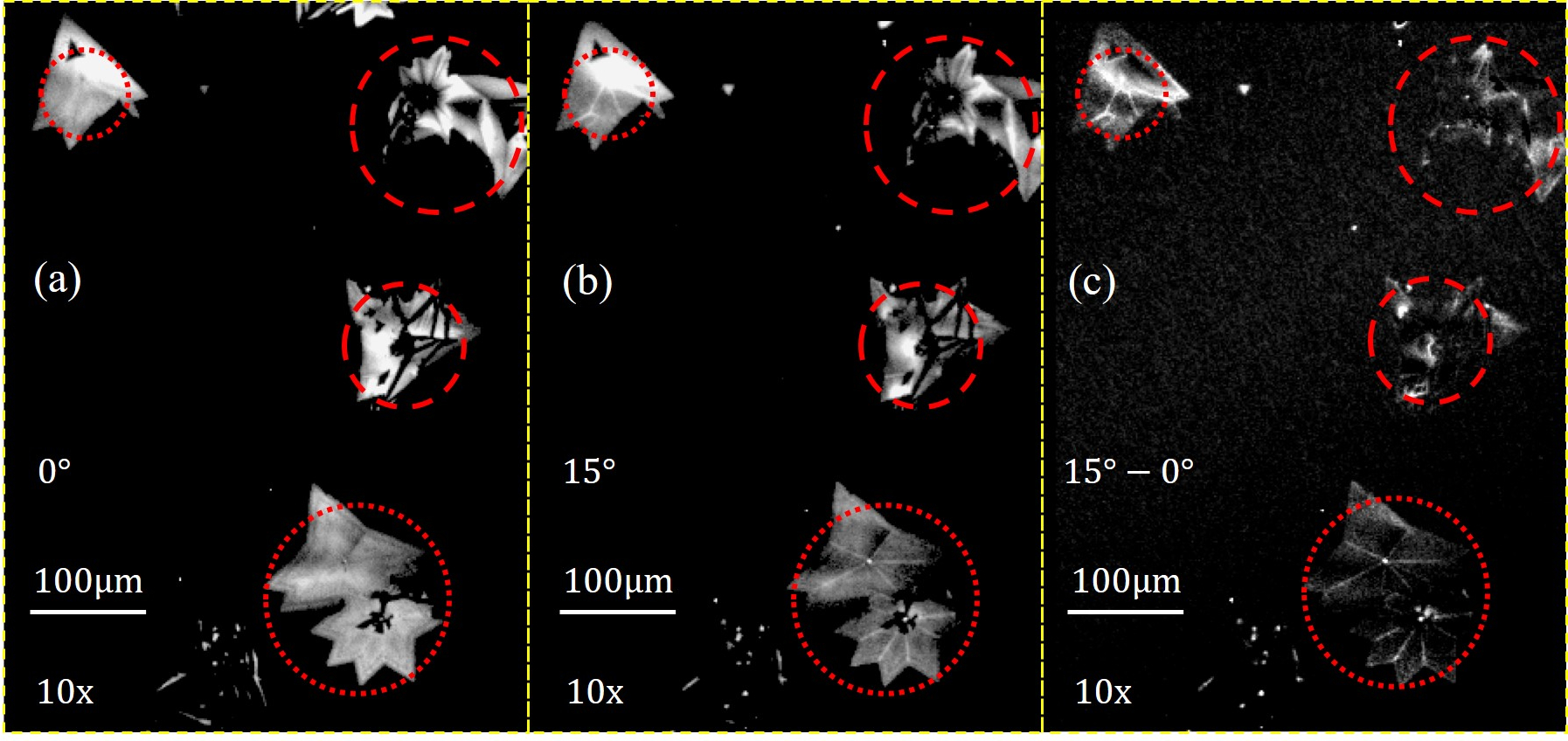}
\caption{Angular dependent PL images of WS$_{2}$ monolayer flake, excited by 450 nm laser, imaged by a 10x objective lens, showing multiple flakes in the field of view that respond to rotations of the bandpass filter. Red circles highlight flakes that showed the most pronounced changes. (a) $0^{\circ}$ , (b) $15^{\circ}$ and (c) intensity differences between $0^{\circ}$ and $15^{\circ}$.}
\label{Fig4}
\end{figure}
Other areas on this sample also displayed similar filter angle dependent photoluminescence. We once again show the images taken with the 10x lens in Figure \ref{Fig4}, except with a larger image area. In this case we highlight three other emissive areas in the vicinity of the one previously investigated. Structures similar to the grain boundaries identified before exhibit blueshifted photoluminescence detected using only higher BPF angles ($\theta=15.0^{\circ},\lambda=626.1$ $nm$, $\Delta\omega=4.1$ $nm$). We identify two flakes containing line defects that behaved in this way. In contrast, the image signal detected from areas on the domains away from the ``lines'' attenuate with increasing $\theta$.

Atomic force microscopy (AFM) was used to investigate the nanoscale origin of the features highlighted in the optical measurement in Figure \ref{Fig3}. The observed behavior of the optical emission, originating from the flake vertices and grain boundaries during angular dependence PL meant that the topographical structure of these areas was of particular interest. AFM images were taken with a Bruker MultiMode 8 scanning probe microscope in Tapping Mode using Budget Sensors Tap300Al-G probes.
\begin{figure}[H]
\centering
\includegraphics[width = 150mm]{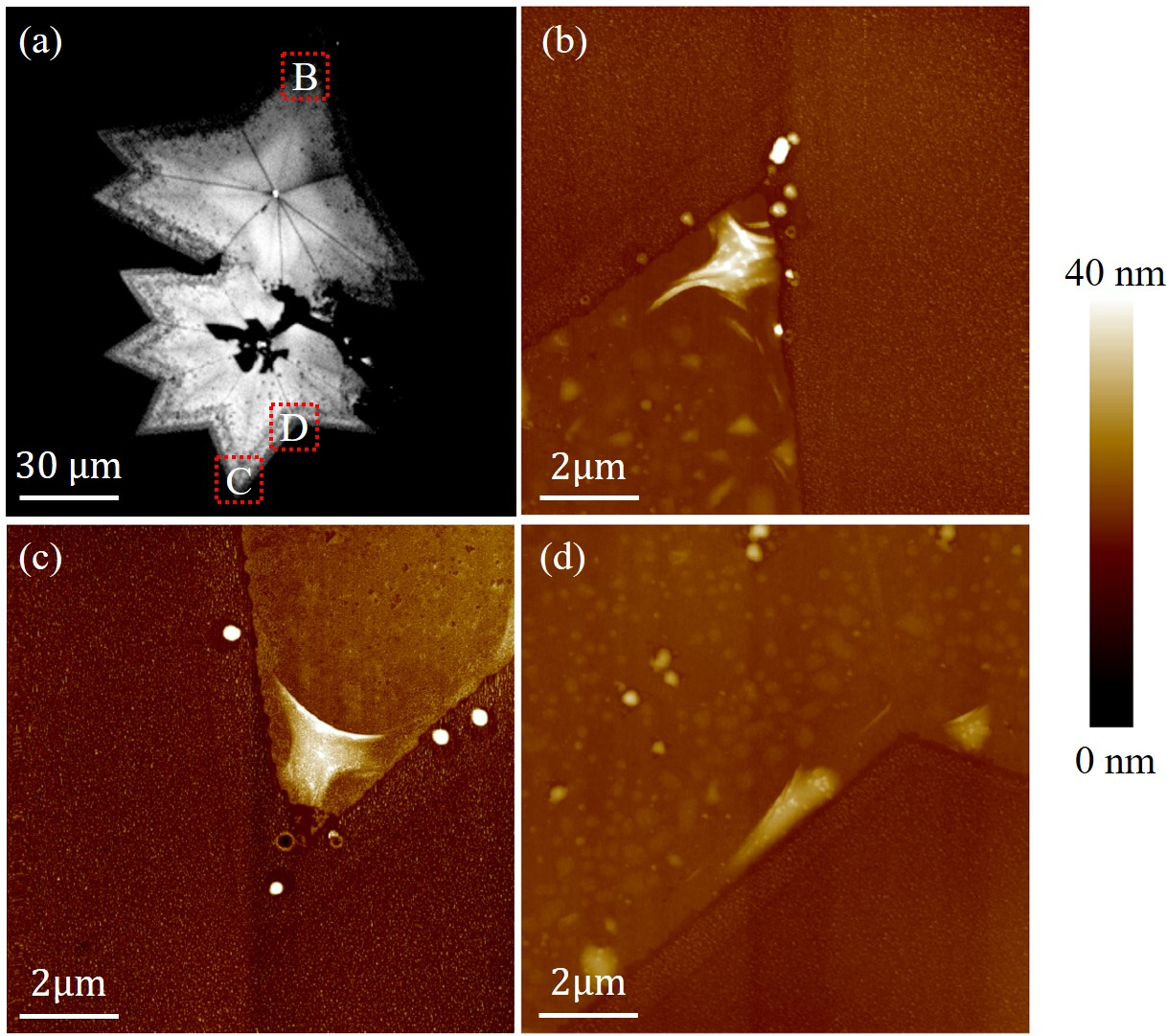}
\caption{Figure 5 Measurements of vertices on WS$_{2}$ Flake 1 in tapping mode AFM in air. Scan area $10$ $\mu$m$\times10$ $\mu$m}
\label{Fig5}
\end{figure}
A large number of defects on the flake surface were found, confirming the BPF angular dependent photoluminescence previously discussed, in the form of either holes or protruding features. Whereas the majority of the flake is monolayer in thickness, the end of each outer vertex, shown in Figure \ref{Fig5}(b) and (c), is characterized by the presence of a much larger structure approximately 25 nm in height and 1.5 $\mu$m across. Similar features up to 10 nm in height also populate the flake edge. The presence of these structures is mainly observed to be within 1 $\mu$m of the flake edge, with few large scale features being observed across the majority of the flake surface. Grain boundaries between the inner vertices, shown in \ref{Fig5}(d), along with the center of the flake can also be observed and appear to be bi-layer in thickness. These grain boundaries correspond to regions in angular dependence PL measurements that peak in intensity towards the blue end of the wavelength range.
\section{Discussion}
The key to implementing a real authentication or identification system based on an optical PUF such as the one we described, is to capture the random response or physical characteristics and generate unique fingerprints. So far we have discussed a method to capture the response from 2D materials, based on imaging the fluorescence from structural defects with a bandpass filter at different detection angles. By adding a bandpass filter, we convolute the spatially varying PL of a given monolayer sample with the transmission response of the bandpass filter at different angles, the difficulty of determining each of these two contributions from the resulting convolution determine the overall complexity and therefore strength of the PUF.

The spatially varying PL of 2D flakes is a result of point, line and other structural defects in the crystal as well as dopants, regardless of the method of synthesis. In fact we showed that both mechanical exfoliation and chemical vapor deposition generate flakes exhibiting spatially varying PL. For mechanically exfoliated flakes, edges and vertices inevitably vary not only in their atomic and chemical properties but also in orientation, because the shear transfer from the tape to a single atomic sheet is dependent on uncontrollable crystal lattice dislocations. On the other hand, the spatial homogeneity of flakes synthesized using chemical vapor transport largely depend on nucleation temperature and growth time. It has been demonstrated that long growth time and high nucleation temperature can lead to flakes with high degrees spatial homogeneity, where the distribution of point defects may be controlled by changing the proportion of reactants. In contrast, shorter growth times and lower reactor temperatures reduces the control over spatial homogeneity, producing a large number of defects of varying types, whose distribution maybe based on stochastic processes \cite{GovindRajan2016}. The identification of the chemistry of defects requires nuclear magnetic resonance, a complex and expensive procedure. Yet to gain a complete structural map for cloning, all kinds of defects have to be identified and understood exactly on each and every flake. Even then, the dynamic surface energy distribution during growth implies an inherent obstacle to obtain exact geometric replicas. It is sufficient for the context of PUFs, that in the absence of knowledge on the precise atomic and chemical makeup of each monolayer it is not possible to make an exact copy of flakes exhibiting identical optical response.

Images in Figure \ref{Fig3}-\ref{Fig4} were obtained with minute long integration times on a thermoelectrically cooled CCD. This was necessary to improve the signal-to-noise ratio to an acceptable level, for our image processing algorithms. At ambient conditions, the signal-to-noise ratio in our system is limited by the low extraction efficiency. This raw efficiency can be improved with e.g., superacid treatment \cite{Alharbi2017}. However for an integrated optical authentication system it may also be necessary to apply an ultrathin layer of covering polymer, such as polymethylmetacrylate (PMMA), that protects the flake from optical degradations as well as damage. It has been proposed that photonic crystal cavities could be engineered for 2D materials to enable Purcell enhancement via optimized spatial and spectral coupling to cavity modes, leading to improved $\eta_{q}$ \cite{Noori2016}. Micro-lenses such as UV curable epoxy solid immersion lenses can also be integrated to protect the flakes as well as further optimizing the extracting efficiency \cite{Woodhead2016}.

\section{Conclusion}
We report a novel method to implement optically variable physical unclonable functions, using 2D emitters and an angle variable transmission filter. Atomic scale defects have an apparently random distribution that leads to optically measurable unique signatures, comprising of detection wavelength and spatially dependent photoluminescence. It was shown that an inexpensive bandpass filter was sufficient, in terms of transmission variability with angle, to select a range of unique optical patterns from a monolayer WS$_{2}$. For $\theta$ close to normal incidence, monolayer emission dominates the optical image and for more oblique angles contribution from line and point defects take precedence. Similar variations were found for other flakes on the same sample, hence for each BPF angle we can select a unique optical signature based on spatially varying PL contributions from each and every flake on the sample. Crystal nucleation and growth generate a continuous spectrum of different geometries even for identical growth conditions, improving the strength of our 2D material OPUF by making it manufacturer resistant. Spatial non-uniform photoluminescence is more pronounced for chemical vapor grown flakes than those created using mechanical exfoliation, as confirmed by AFM that showed a rich distribution of structural defects.
Finally, we identify basic security considerations as well as suggestions to improve the detection efficiency of optical signatures. This work paves the way to implementing robust authentication systems protected from cloning at the atomic level.
\section{Acknowledgements}
RJY and VIF conceived the experiment. Chemical vapor deposition growth was carried out by AA and DS who also optimized the samples for photoluminescence. YC made the mechanical exfoliated sample. YC and AJR setup the experiment and carried out the measurements. YC analysed the data. JR, CSW, YJN, RB-G, UR helped design the experiment. The manuscript was prepared primarily by YC and RJY with contribution from all authors.

RJY acknowledges support by the Royal Society through a University Research Fellowship (UF110555 and UF160721). DS and AA acknowledges support by the National Science Foundation award (\#1638598). This material is based upon work supported by the Air Force Office of Scientific Research under award number FA9550-16-1-0276. This work was also supported by grants from The Engineering and Physical Sciences Research Council in the UK (EP/K50421X/1 and EP/L01548X/1).

\end{document}